\documentclass[nofootinbib,twocolumn,preprintnumbers,superscriptaddress]{revtex4-1}
\pdfoutput=1

\RequirePackage{snapshot} 
\usepackage{amsmath}
\usepackage{amssymb}
\usepackage{graphicx}
\usepackage{hyperref}
\usepackage{xspace}
\usepackage{wasysym} 

\usepackage{booktabs}
\AtBeginDocument{
\heavyrulewidth=.08em
\lightrulewidth=.05em
\cmidrulewidth=.03em
\belowrulesep=.65ex
\belowbottomsep=0pt
\aboverulesep=.4ex
\abovetopsep=0pt
\cmidrulesep=\doublerulesep
\cmidrulekern=.5em
\defaultaddspace=.5em
}

\newcommand{\as}{\alpha_s}
\newcommand{\muR}{\mu_R}
\newcommand{\muF}{\mu_F}

\newcommand{\GeV}{\;\mathrm{GeV}}
\newcommand{\TeV}{\;\mathrm{TeV}}

\newcommand{\NNLO}{\text{NNLO}}
\newcommand{\NNNLO}{\text{N$^3$LO}}
\usepackage{xcolor}
\definecolor{light-gray}{gray}{0.8}

\begin{document}
\title{Vector-boson fusion Higgs production at \NNNLO{} in QCD}

\preprint{CERN-TH-2016-127}
\preprint{OUTP-16-12P}

\newcommand{\CERNaff}{CERN, Theoretical Physics Department, 
  CH-1211 Geneva 23, Switzerland}
\newcommand{\CNRSaff}{CNRS, UMR 7589, LPTHE, F-75005, Paris, France}
\newcommand{\SORBaff}{Sorbonne Universit\'es, UPMC Univ Paris 06, UMR
  7589, LPTHE, F-75005, Paris, France}
\newcommand{\OXFaff}{Rudolf Peierls Centre for Theoretical Physics,
  1 Keble Road, University of Oxford, UK}
\author{Fr\'ed\'eric A. Dreyer}
\affiliation{\SORBaff}
\affiliation{\CNRSaff}
\affiliation{\CERNaff}
\author{Alexander Karlberg}
\affiliation{\OXFaff}

\begin{abstract} 
  We calculate the next-to-next-to-next-to-leading-order (\NNNLO{})
  QCD corrections to inclusive vector-boson fusion (VBF) Higgs production at
  proton colliders, in the limit in which there is no colour exchange
  between the hadronic systems associated with the two colliding
  protons.
  We also provide differential cross sections for the Higgs transverse
  momentum and rapidity distributions.
  We find that the corrections are at the $1-2\permil$ level, well
  within the scale-uncertainty of the next-to-next-to-leading-order (NNLO) calculation.
  The associated scale-uncertainty of the \NNNLO{} calculation is
  typically found to be below the $2\permil$ level.
  We also consider theoretical uncertainties due to missing higher
  order parton distribution functions, and provide an estimate of
  their importance.
\end{abstract}

\maketitle
\begin{figure}[ht]
  \centering
  \includegraphics[width=0.75\linewidth]{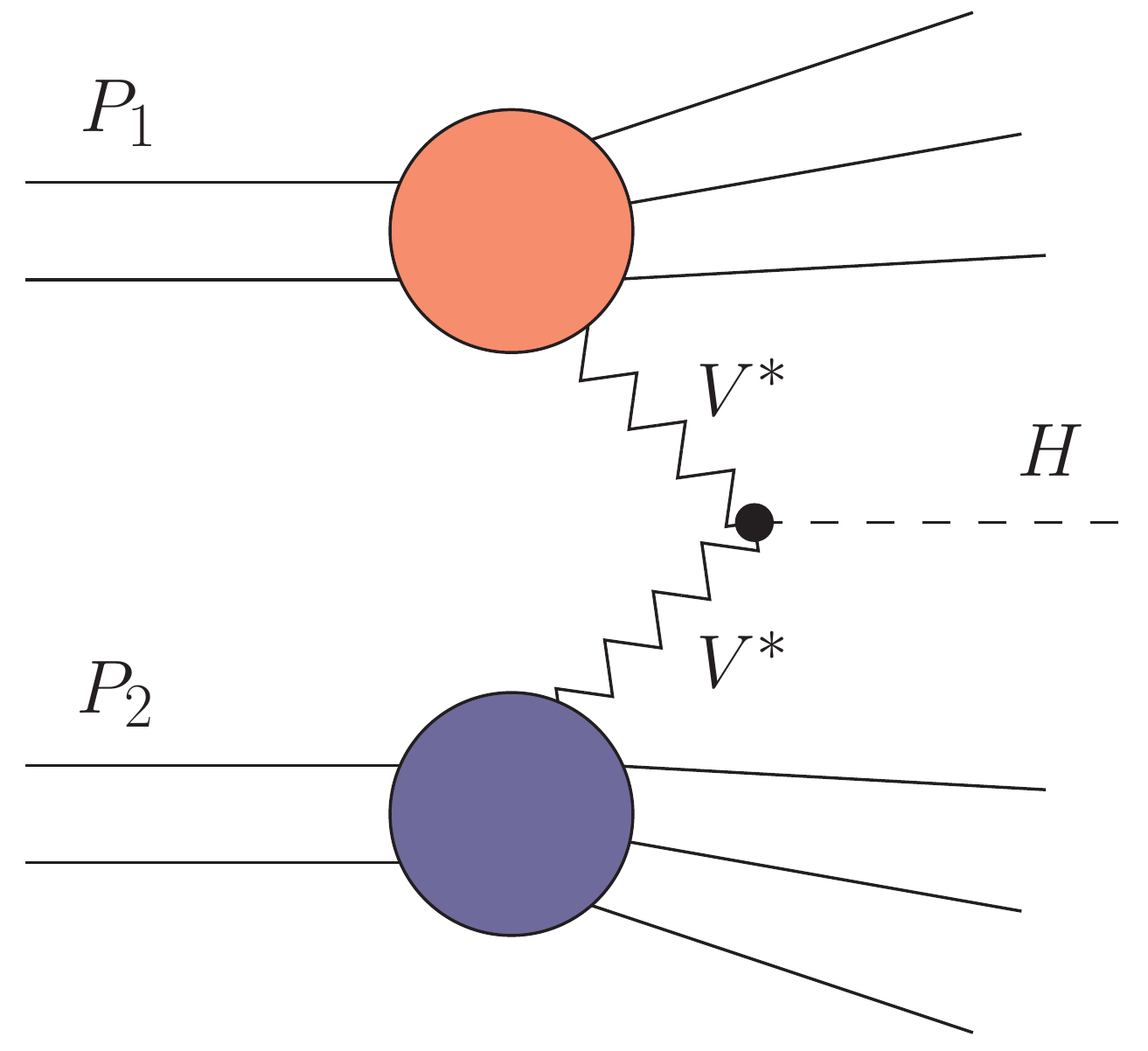}
  \caption{Illustration of Higgs production through vector-boson fusion.}
  \label{fig:vbfh}
\end{figure}

Since the discovery of the Higgs boson
\cite{Aad:2012tfa,Chatrchyan:2012ufa}, the LHC has commenced a program
of precision studies of its properties.
Higgs production through vector-boson fusion (VBF), shown in
figure~\ref{fig:vbfh}, is a key process for precision measurements of
properties of the Higgs
boson~\cite{Khachatryan:2015bnx}, as it is a clean
channel with very distinctive kinematics, due to its $t$-channel
production and to the presence of two high rapidity jets in the final
state.
These features provide an ideal access for the intricate measurements
of the Higgs couplings~\cite{Zeppenfeld:2000td}.
Currently the VBF production signal strength has been measured with a
precision of about 24\%~\cite{ATLASCMS:2015higgs}, though significant
improvements can be expected during run 2 and with the high luminosity
LHC.

In order to experimentally determine the properties of the Higgs boson it is
crucial to have very precise theoretical predictions for cross sections. 
The inclusive cross section for VBF Higgs production is known to
next-to-next-to-leading order (NNLO)~\cite{Bolzoni:2010xr,Bolzoni:2011cu} in the
structure function approach, in which VBF-induced Higgs production is treated as
a double deep-inelastic scattering (DIS) process~\cite{Han:1992hr}.
This calculation found NNLO corrections of about $1\%$ and
renormalisation and factorisation scale uncertainties at the
$5\permil$ level.
Recently, the fully differential NNLO QCD corrections in VBF Higgs production
have been computed \cite{Cacciari:2015jma}.
These where found to be significant after typical VBF cuts, with
corrections up to 10--12\% in certain kinematical regions.
The calculation also showed no significant reduction in the associated scale
uncertainties compared to the scale uncertainty at next-to-leading-order (NLO).

The structure function approximation is known to be very accurate for VBF,
because non-factorisable colour exchanges are both kinematically and colour
suppressed, such that they are expected to contribute to less than $1\%$ of the
cross section~\cite{Ciccolini:2007ec,Harlander:2008xn,Bolzoni:2011cu}.
This approach is exact in the limit in which one considers that there
are two identical copies of QCD associated with each of the two
protons (shown orange and blue in figure~\ref{fig:vbfh}), whose
interaction is mediated by the weak force.

In this letter we compute the next-to-next-to-next-to-leading order (\NNNLO{})
QCD corrections to the inclusive cross section in the structure function
approach.
This calculation provides the second \NNNLO{} calculation for
processes of relevance to the LHC physics program, after a similar
accuracy was recently achieved in the gluon-gluon fusion
channel~\cite{Anastasiou:2015ema}.
It represents an important milestone towards achieving a fully
differential \NNNLO{} calculation with the projection-to-Born
method~\cite{Cacciari:2015jma}.
We also provide an estimate of contributions to the cross section from
missing higher order parton distribution functions (PDFs) as these are
currently only known at NNLO.

In the structure function approach the VBF Higgs production cross
section is calculated as a double DIS process and can thus be
expressed as~\cite{Han:1992hr}
\begin{align}
  \label{eq:vbfh-dsigma}
  d\sigma = &\frac{4\sqrt{2}}{s} G_F^3 m^8_{V}
  \Delta_V^2(Q_1^2)
  \Delta_V^2(Q_2^2) \notag
  \\
  &\times
  \mathcal{W}^V_{\mu\nu}(x_1,Q_1^2) 
  \mathcal{W}^{V,\mu\nu}(x_2,Q_2^2)
  d\Omega_{\text{VBF}} \,.
\end{align}
Here $G_F$ is Fermi's constant, $m_V$ is the mass of the vector boson,
$\sqrt{s}$ is the collider centre-of-mass energy, $\Delta_V^2$ is the squared
boson propagator, $Q_i^2 = -q_i^2$ and $x_i = Q_i^2/(2P_i\cdot q_i)$ are the
usual DIS variables, and $d\Omega_{\text{VBF}}$ is the three-particle VBF phase
space.
The hadronic tensor $\mathcal{W}^V_{\mu\nu}$ can be expressed as
\begin{multline}
  \label{eq:hadr-tensor}
  \mathcal{W}^V_{\mu\nu}(x_i,Q_i^2) = 
  \Big(-g_{\mu\nu}+\frac{q_{i,\mu}q_{i,\nu}}{q_i^2}\Big) F_1^V(x_i,Q_i^2)
  \\
  + \frac{\hat{P}_{i,\mu}\hat{P}_{i,\nu}}{P_i\cdot q_i} F_2^V(x_i,Q_i^2)
  + i\epsilon_{\mu\nu\rho\sigma}\frac{P_i^\rho q_i^\sigma}{2 P_i\cdot q_i} 
  F_3^V(x_i,Q_i^2)\,,
\end{multline}
where we defined
$\hat{P}_{i,\mu} = P_{i,\mu} - \tfrac{P_i \cdot q_i}{q_i^2}
q_{i,\mu}$, and the $F^V_i(x,Q^2)$ functions are the standard DIS
structure functions with $i=1,2,3$ and $V=Z,W^-,W^+$.

From the knowledge of the vector-boson momenta $q_i$, it is
straightforward to reconstruct the Higgs momentum.
As such, the cross section obtained using
equation~(\ref{eq:vbfh-dsigma}) is differential in the Higgs
kinematics.

In order to compute the N$^n$LO cross section, we require the structure
functions $F_i^V$ up to order $\mathcal{O}(\alpha_s^n)$ in the strong
coupling constant.
We express the structure functions as convolutions of the PDFs with
the short distance coefficient functions
\begin{equation}
  \label{eq:conv-structf}
  F_i^V = \sum_{a=q,g} C_i^{V,a} \otimes f_a \,,\quad i=1,2,3\,.
\end{equation}
All the necessary coefficient functions are known up to third order.%
\footnote{The even-odd differences between charged-current coefficient
  functions are known only approximately, since only the five lowest
  moments have been calculated~\cite{Moch:2007rq}. 
  However, the uncertainty associated with this approximation is less than
  $1\permil$ of the \NNNLO{} correction, and therefore completely negligible.}
To compute the \NNNLO{} VBF Higgs production cross section, one can
therefore evaluate the convolution of the PDF with the appropriate
coefficient functions in equation~(\ref{eq:conv-structf}).
At \NNNLO{}, additional care is required due to the appearance of new
flavour topologies~\cite{Vermaseren:2005qc}.
As such, contributions corresponding to interferences of diagrams where the
vector boson attaches on different quark lines are to be set to zero for
charged boson exchanges.

To compute the dependence of the cross section on the values of the
factorisation and renormalisation scales, we use renormalisation group
methods~\cite{Furmanski:1981cw,vanNeerven:2000uj,Buehler:2013fha}, and
evaluate the scale dependence to third order in the coefficient
functions as well as in the PDFs.
The running of the coefficient functions can be obtained using the
first two terms in the expansion of the beta function.
To obtain the dependence of the PDFs on the factorisation scale, we
integrate the parton density evolution equation.
For completeness, the technical details of this procedure are given in
the supplemental material of this letter~\cite{FDAKadditional}.

There is one source of formally \NNNLO{} QCD corrections appearing in
equation~(\ref{eq:conv-structf}) which is currently unknown, namely
missing higher order terms in the determination of the PDF.
Indeed, while one would ideally calculate the \NNNLO{} cross section
using \NNNLO{} parton densities, only NNLO PDF sets are available at
this time.
These will be missing contributions from two main sources: from the
higher order corrections to the coefficient functions that relate
physical observables to PDFs; and from the higher order splitting
functions in the evolution of the PDFs.

To evaluate the impact of future \NNNLO{} PDF sets on the total cross
section, we consider two different approaches.
A first, more conservative estimate, is to derive the uncertainty
related to higher order PDF sets from the difference at lower orders,
as described in~\cite{Anastasiou:2016cez} (see
also~\cite{Forte:2013mda}).
We compute the NNLO cross section using both the NLO and the NNLO PDF
set, and use their difference to extract the \NNNLO{} PDF uncertainty.
We find in this way that at 13 TeV the uncertainty from missing higher
orders in the extractions of PDFs is
\begin{equation}
  \label{eq:pdf-uncert-schemeA}
  \delta_A^{\text{PDF}} = \frac12
  \left|\frac{\sigma^\text{NNLO}_\text{NNLO-PDF} - \sigma^\text{NNLO}_\text{NLO-PDF}}{\sigma^\text{NNLO}_\text{NNLO-PDF}}\right|
  = 1.1 \%\,.
\end{equation}
Because the convergence is greatly improved going from NNLO to \NNNLO{} compared
to one order lower, one might expect this to be rather conservative
even with the factor half in equation~(\ref{eq:pdf-uncert-schemeA}).
Therefore, we also provide an alternative estimate of the impact of
higher orders PDFs, using the known \NNNLO{} $F_2$ structure function.

We start by rescaling all the parton distributions using the $F_2$
structure function evaluated at a low scale $Q_0$.
\begin{equation}
  \label{eq:n3lo-pdf-approx}
  f^{\NNNLO, \text{approx.}}(x, Q) = 
  f^{\NNLO}(x,Q) \frac{F_2^{\NNLO}(x, Q_0)}{F_2^{\NNNLO}(x, Q_0)}\,.
\end{equation}
In practice, we will use the $Z$ structure function.
We then re-evaluate the structure functions in
equation~(\ref{eq:conv-structf}) using the approximate higher order
PDF given by equation~(\ref{eq:n3lo-pdf-approx}).
This yields
\begin{equation}
  \label{eq:pdf-uncert-schemeB}
  \delta_B^{\text{PDF}}(Q_0) = \left|\frac{\sigma^\NNNLO - \sigma^\NNNLO_\text{rescaled}(Q_0)}{\sigma^\NNNLO}\right|
  = 7.9 \text{\permil}\,,
\end{equation}
where in the last step, we used $Q_0=8\GeV$ and considered $13\TeV$
proton collisions.

By calculating a rescaled NLO PDF and evaluating the NNLO cross
section in this way, we can evaluate the ability of this method to
predict the corrections from NNLO PDFs.
We find that with $Q_0=8\GeV$, the uncertainty estimate obtained in
this way captures relatively well the impact of NNLO PDF sets.

The rescaled PDF sets obtained using
equation~(\ref{eq:n3lo-pdf-approx}) will be missing \NNNLO{}
corrections from the evolution of the PDFs in energy.
We have checked the impact of these terms by varying the
renormalisation scale up and down by a factor two around the
factorisation scale in the splitting functions used for the PDF
evolution.
We find that the theoretical uncertainty associated with missing
higher order splitting functions is less than one permille of the
total cross section.
Comparing this with equation~(\ref{eq:pdf-uncert-schemeB}), it is
clear that these effects are numerically subleading, suggesting that a
practical alternative to full \NNNLO{} PDF sets could be obtained by
carrying out a fit of DIS data using the hard \NNNLO{} matrix element.
We leave a detailed study of this question for future work.

\begin{figure}
  \centering
  \includegraphics[width=0.98\linewidth]{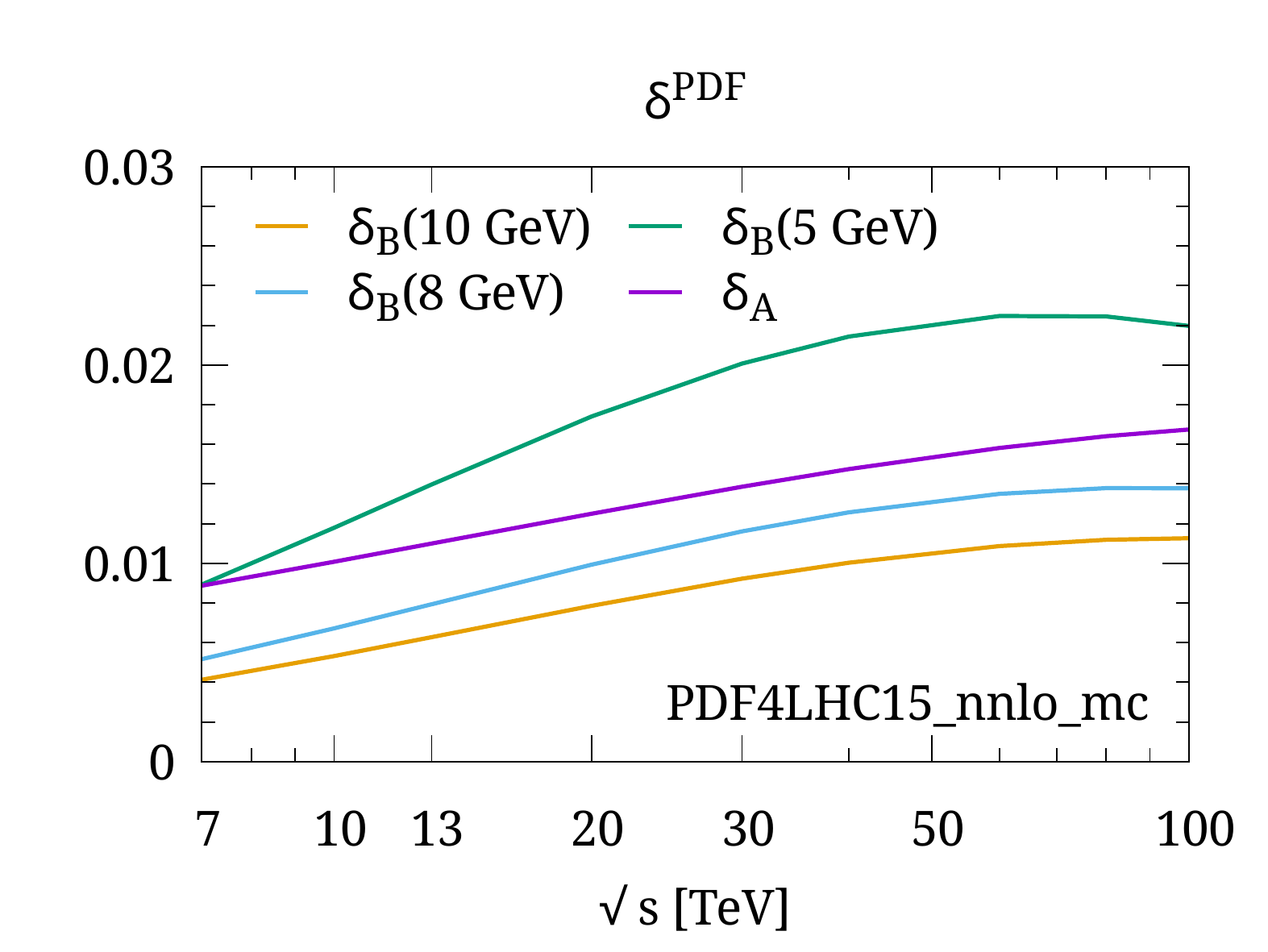}
  \caption{Estimate of the impact of missing higher orders corrections in PDFs, 
    using equations~(\ref{eq:pdf-uncert-schemeA}) 
    and~(\ref{eq:pdf-uncert-schemeB}) with $Q_0=5$, $8$ and $10\GeV$.
    \label{fig:n3lo-pdf}
  }
\end{figure}

The uncertainty estimates obtained with the two different methods
described by equations~(\ref{eq:pdf-uncert-schemeA})
and~(\ref{eq:pdf-uncert-schemeB}) is shown in
figure~\ref{fig:n3lo-pdf} as a function of center-of-mass energy, and
for a range of $Q_0$ values.

One should note that the uncertainty estimates given in
equations~(\ref{eq:pdf-uncert-schemeA})
and~(\ref{eq:pdf-uncert-schemeB}) do not include what is usually
referred to as PDF uncertainties.
While we are here calculating missing higher order uncertainties to
NNLO PDF sets, typical PDF uncertainties correspond to uncertainties
due to errors on the experimental data and limitations of the fitting
procedure.
These can be evaluated for example with the PDF4LHC15
prescription~\cite{Butterworth:2015oua}, and are of about $2\%$ at
$13\TeV$, which is larger than the corrections discussed above.
One can also combine them with $\as$ uncertainties, which are at the
$5\permil$ level.

Let us now discuss in more detail phenomenological consequences of the
\NNNLO{} corrections to VBF Higgs production. 
We present results for a wide range of energies in proton-proton
collisions.
The central factorisation and renormalisation scales are set to the
squared momentum of the corresponding vector boson.
To estimate missing higher-order uncertainties, we use a seven-point
scale variation, varying the scales by a factor two up and down
while keeping $0.5 < \mu_R/ \mu_F < 2$
\begin{equation}
  \label{eq:scale-choice}
  \mu_{R,i} = \xi_{\mu_{R}} Q_i\,,\quad
  \mu_{F,i} = \xi_{\mu_{F}} Q_i\,,
\end{equation}
where $\xi_{\mu_R},\xi_{\mu_F}\in\big\{\tfrac12,1,2\big\}$ and $i=1$, $2$
corresponds to the upper and lower hadronic sectors.

Our implementation of the calculation is based on the inclusive part of
\texttt{proVBFH} which was originally developed for the differential NNLO VBF
calculation~\cite{Cacciari:2015jma}. We have used the phase space from
\texttt{POWHEG}'s two-jet VBF Higgs calculation~\cite{Nason:2009ai}.
The matrix element is derived from structure functions obtained with the
parametrised DIS coefficient
functions~\cite{SanchezGuillen:1990iq,vanNeerven:1991nn,Zijlstra:1992qd,Zijlstra:1992kj,vanNeerven:1999ca,vanNeerven:2000uj,Moch:2004xu,Vermaseren:2005qc,Vogt:2006bt,Moch:2007rq},
evaluated using \texttt{HOPPET} v1.2.0-devel~\cite{Salam:2008qg}.

For our computational setup, we use a diagonal CKM matrix with five light
flavours ignoring top-quarks in the internal lines and final states. Full
Breit-Wigner propagators for the $W$, $Z$ and the narrow-width approximation for
the Higgs boson are applied.
We use the PDF4LHC15\_nnlo\_mc
PDF~\cite{Butterworth:2015oua,Dulat:2015mca,Harland-Lang:2014zoa,Ball:2014uwa}
and four-loop evolution of the strong coupling, taking as our initial condition
$\as(M_Z) = 0.118$.
We set the Higgs mass to $M_H = 125.09\GeV$, in accordance with the
experimentally measured value~\cite{Aad:2015zhl}.  Electroweak
parameters are obtained from their PDG~\cite{Agashe:2014kda} values
and tree-level electroweak relations. As inputs we use
$M_W = 80.385\GeV$, $M_Z = 91.1876\GeV$ and
$G_F = 1.16637\times 10^{-5} \GeV^{-2}$. For the widths of the vector
bosons we use $\Gamma_W = 2.085 \GeV $ and $\Gamma_Z = 2.4952 \GeV$.

\begin{figure}
  \centering
  \includegraphics[page=1,width=0.98\linewidth]{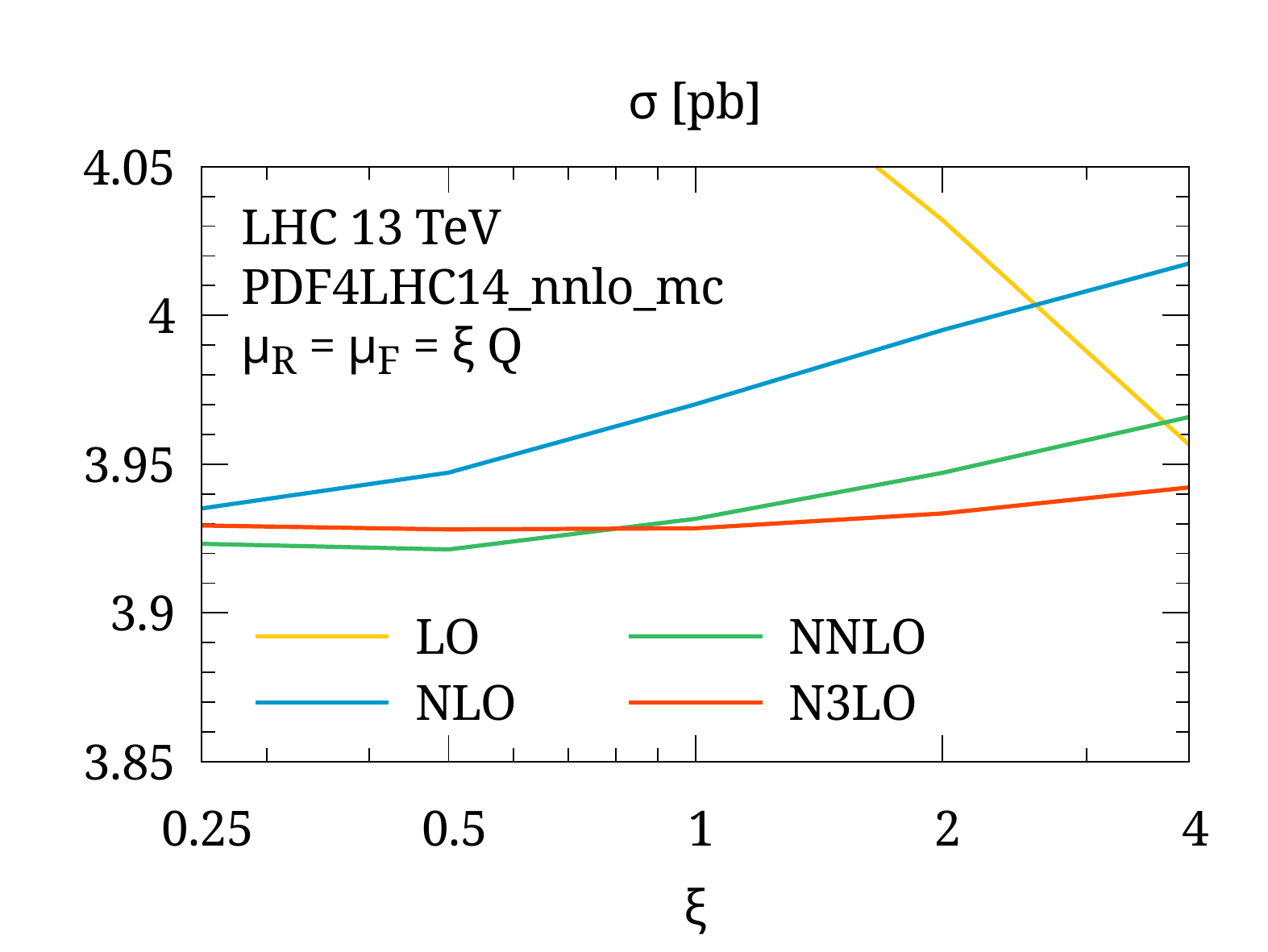}%
  \caption{Dependence of the cross section on the renormalisation 
    and factorisation scales for each order in perturbation theory.
    \label{fig:scale-var} }
\end{figure}

To study the convergence of the perturbative series, we show in
figure~\ref{fig:scale-var} the inclusive cross section obtained at 13
TeV with $\mu_R=\mu_F=\xi Q$ for $\xi\in [1/4,4]$.
Here we observe that at \NNNLO{} the scale dependence becomes
extremely flat over the full range of renormalisation and
factorisation scales.
We note that similarly to the results obtained in the the gluon-fusion
channel~\cite{Anastasiou:2015ema}, the convergence improves
significantly at \NNNLO{}, with the \NNNLO{} prediction being well
inside of the NNLO uncertainty band, while at lower orders there is a
pattern of limited overlap of theoretical uncertainties.
\begin{figure*}[th]
  \centering
  \includegraphics[page=1,width=0.35\textwidth]{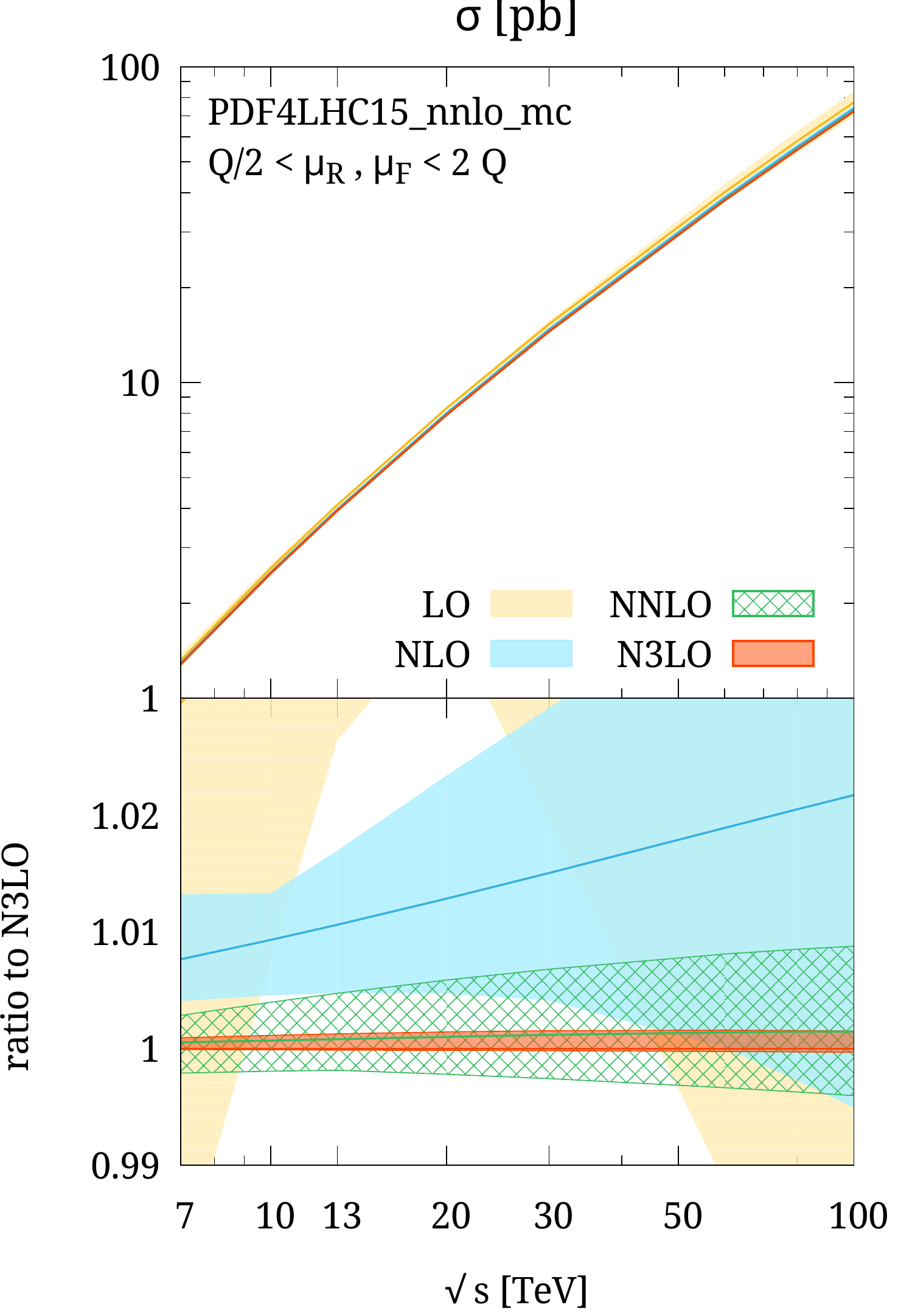}%
  \hspace{-5mm}%
  \hfill\includegraphics[page=1,width=0.35\textwidth]{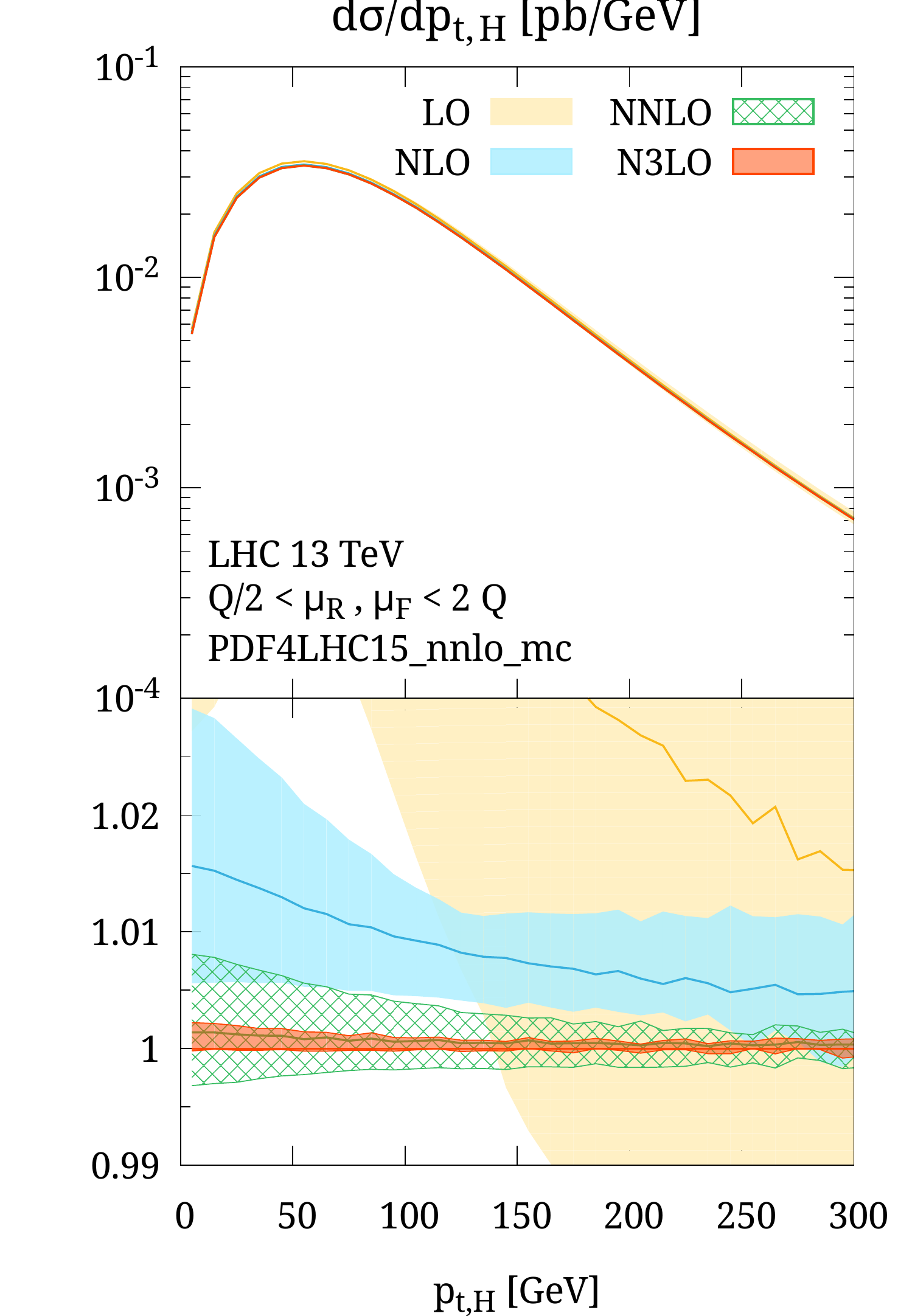}%
  \hspace{-5mm}%
  \hfill\includegraphics[page=2,width=0.35\textwidth]{n3lo-hists}
  \caption{Cross section as a function of center-of-mass energy
    (left), Higgs transverse momentum distribution (center) and Higgs
    rapidity distribution (right).
    \label{fig:distributions} }
\end{figure*}

In figure~\ref{fig:distributions} (left), we give the cross section as
a function of center-of-mass energy.
We see that at \NNNLO{} the convergence of the perturbative series
is very stable, with corrections of about $1\permil$ on the NNLO
result.
The scale uncertainty is dramatically reduced, going at $13 \TeV$ from
$7\permil$ at NNLO to $1.4\permil$ at \NNNLO.
A detailed breakdown of the cross section and scale uncertainty
obtained at each order in QCD is given in
table~\ref{tab:cross-sections} for $\sqrt{s}=13$, $14$ and $100\TeV$.

The center and right plots of figure~\ref{fig:distributions} show the
Higgs transverse momentum and rapidity distributions at each order in
QCD, where we observe again a large reduction of the theoretical
uncertainty at \NNNLO{}.

\begin{table}[t]
  \centering
  \phantom{x}\medskip
  \begin{tabular}{lcccccc}
    \toprule
    &&  $\sigma^{(13 \TeV)}$  [pb]  && $\sigma^{(14 \TeV)}$ [pb]  && $\sigma^{(100 \TeV)}$ [pb] \\
    \midrule
    LO       &&  $4.099\,^{+0.051}_{-0.067}$    &&  $4.647\,^{+0.037}_{-0.058}$   
             &&  $77.17\,^{+6.45}_{-7.29}$ \\[4pt]
    NLO      &&  $3.970\,^{+0.025}_{-0.023}$    &&  $4.497\,^{+0.032}_{-0.027}$   
             &&  $73.90\,^{+1.73}_{-1.94}$ \\[4pt]
    NNLO     &&  $3.932\,^{+0.015}_{-0.010}$    &&  $4.452\,^{+0.018}_{-0.012}$
             &&  $72.44\,^{+0.53}_{-0.40}$ \\[4pt]
    \NNNLO{} &&  $3.928\,^{+0.005}_{-0.001}$    &&  $4.448\,^{+0.006}_{-0.001}$ 
             &&  $72.34\,^{+0.11}_{-0.02}$ \\
    \bottomrule
  \end{tabular}
  \caption{Inclusive cross sections at LO, NLO, NNLO and \NNNLO{} for VBF Higgs production.
    The quoted uncertainties correspond to scale variations $Q/2 < \muR, \muF < 2 Q$,
    while statistical uncertainties are at the level of $0.2\permil$.
    \label{tab:cross-sections}}
\end{table}

A comment is due on non-factorisable QCD corrections.
Indeed, for the results presented in this letter, we have considered
VBF in the usual DIS picture, ignoring diagrams that are not of the
type shown in figure~\ref{fig:vbfh}.
These effects neglected by the structure function approximation are
known to contribute less than $1\%$ to the total cross section at NNLO
\cite{Bolzoni:2011cu}.
The effects and their relative corrections are as follows:
\begin{itemize}
\item Gluon exchanges between the upper and lower ha\-dro\-nic
  sectors, which appear at NNLO, but are kinematically and colour
  suppressed. 
  These contributions along with the heavy-quark loop induced
  contributions have been estimated to contribute at the permille
  level~\cite{Bolzoni:2011cu}.
  
\item t-/u-channel interferences which are known to contribute
  $\mathcal{O}(5\permil)$ at the fully inclusive level and
  $\mathcal{O}(0.5\permil)$ after VBF cuts have been applied
  \cite{Ciccolini:2007ec}.
  
\item Contributions from s-channel production, which have been
  calculated up to NLO~\cite{Ciccolini:2007ec}.
  At the inclusive level these contributions are sizeable but they are
  reduced to $\mathcal{O}(5\permil)$ after VBF cuts.

\item Single-quark line contributions, which contribute to the VBF
  cross section at NNLO. 
  At the fully inclusive level these amount to corrections of
  $\mathcal{O}(1\%)$ but are reduced to the permille level after VBF
  cuts have been applied~\cite{Harlander:2008xn}.
  
\item Loop induced interferences between VBF and gluon-fusion Higgs
  production. 
  These contributions have been shown to be much below the permille
  level \cite{Andersen:2007mp}.
 
\end{itemize}

Furthermore, for phenomenological applications, one also needs to
consider NLO electroweak effects~\cite{Ciccolini:2007ec}, which amount
to $\mathcal{O}(5\%)$ of the total cross section.
We leave a detailed study of non-factorisable and electroweak effects
for future work.
The code used for this calculation will be published in the near
future~\cite{proVBFH}.

In this letter, we have presented the first \NNNLO{} calculation of a
$2\to 3$ hadron-collider process, made possible by the DIS-like
factorisation of the process.
This brings the precision of VBF Higgs production to the same formal
accuracy as was recently achieved in the gluon-gluon fusion
channel in the heavy top mass approximation~\cite{Anastasiou:2015ema}.
The \NNNLO{} corrections are found to be tiny, $1-2\permil$, and well
within previous theoretical uncertainties, but they provide a large
reduction of scale uncertainties, by a factor 5.
This calculation also provides the first element towards a
differential \NNNLO{} calculation for VBF Higgs production, which
could be achieved through the projection-to-Born
method~\cite{Cacciari:2015jma} using a NNLO DIS 2+1 jet
calculation~\cite{Gehrmann:2009vu}.

\textbf{Acknowledgments:} 
We are grateful to Gavin Salam for numerous suggestions and comments
over the course of this work, and to Giulia Zanderighi for detailed
comments on the manuscript.
We also thank Andreas Vogt for clarifications on the third order
coefficient functions, and Marco Zaro for providing code allowing
numerical cross checks of the NNLO results.
F.D. is supported by the Labex ILP (reference ANR-10-LABX-63) part of
the Idex SUPER, and received financial state aid managed by the Agence
Nationale de la Recherche, as part of the programme Investissements
d'avenir under the reference ANR-11-IDEX-0004-02.
A. K. is supported by the British Science and Technology Facilities Council and
by the Buckee Scholarship at Merton College. A. K. thanks CERN for hospitality
while part of this work was performed.

\bibliography{n3lo_vbfh}{}

\begin{thebibliography}{40}%
\makeatletter
\providecommand \@ifxundefined [1]{%
 \@ifx{#1\undefined}
}%
\providecommand \@ifnum [1]{%
 \ifnum #1\expandafter \@firstoftwo
 \else \expandafter \@secondoftwo
 \fi
}%
\providecommand \@ifx [1]{%
 \ifx #1\expandafter \@firstoftwo
 \else \expandafter \@secondoftwo
 \fi
}%
\providecommand \natexlab [1]{#1}%
\providecommand \enquote  [1]{``#1''}%
\providecommand \bibnamefont  [1]{#1}%
\providecommand \bibfnamefont [1]{#1}%
\providecommand \citenamefont [1]{#1}%
\providecommand \href@noop [0]{\@secondoftwo}%
\providecommand \href [0]{\begingroup \@sanitize@url \@href}%
\providecommand \@href[1]{\@@startlink{#1}\@@href}%
\providecommand \@@href[1]{\endgroup#1\@@endlink}%
\providecommand \@sanitize@url [0]{\catcode `\\12\catcode `\$12\catcode
  `\&12\catcode `\#12\catcode `\^12\catcode `\_12\catcode `\%12\relax}%
\providecommand \@@startlink[1]{}%
\providecommand \@@endlink[0]{}%
\providecommand \url  [0]{\begingroup\@sanitize@url \@url }%
\providecommand \@url [1]{\endgroup\@href {#1}{\urlprefix }}%
\providecommand \urlprefix  [0]{URL }%
\providecommand \Eprint [0]{\href }%
\providecommand \doibase [0]{http://dx.doi.org/}%
\providecommand \selectlanguage [0]{\@gobble}%
\providecommand \bibinfo  [0]{\@secondoftwo}%
\providecommand \bibfield  [0]{\@secondoftwo}%
\providecommand \translation [1]{[#1]}%
\providecommand \BibitemOpen [0]{}%
\providecommand \bibitemStop [0]{}%
\providecommand \bibitemNoStop [0]{.\EOS\space}%
\providecommand \EOS [0]{\spacefactor3000\relax}%
\providecommand \BibitemShut  [1]{\csname bibitem#1\endcsname}%
\let\auto@bib@innerbib\@empty
\bibitem [{\citenamefont {Aad}\ \emph {et~al.}(2012)\citenamefont {Aad} \emph
  {et~al.}}]{Aad:2012tfa}%
  \BibitemOpen
  \bibfield  {author} {\bibinfo {author} {\bibfnamefont {G.}~\bibnamefont
  {Aad}} \emph {et~al.} (\bibinfo {collaboration} {ATLAS Collaboration}),\
  }\href {\doibase 10.1016/j.physletb.2012.08.020} {\bibfield  {journal}
  {\bibinfo  {journal} {Phys.Lett.}\ }\textbf {\bibinfo {volume} {B716}},\
  \bibinfo {pages} {1} (\bibinfo {year} {2012})},\ \Eprint
  {http://arxiv.org/abs/1207.7214} {arXiv:1207.7214 [hep-ex]} \BibitemShut
  {NoStop}%
\bibitem [{\citenamefont {Chatrchyan}\ \emph {et~al.}(2012)\citenamefont
  {Chatrchyan} \emph {et~al.}}]{Chatrchyan:2012ufa}%
  \BibitemOpen
  \bibfield  {author} {\bibinfo {author} {\bibfnamefont {S.}~\bibnamefont
  {Chatrchyan}} \emph {et~al.} (\bibinfo {collaboration} {CMS Collaboration}),\
  }\href {\doibase 10.1016/j.physletb.2012.08.021} {\bibfield  {journal}
  {\bibinfo  {journal} {Phys.Lett.}\ }\textbf {\bibinfo {volume} {B716}},\
  \bibinfo {pages} {30} (\bibinfo {year} {2012})},\ \Eprint
  {http://arxiv.org/abs/1207.7235} {arXiv:1207.7235 [hep-ex]} \BibitemShut
  {NoStop}%
\bibitem [{\citenamefont {Khachatryan}\ \emph {et~al.}(2015)\citenamefont
  {Khachatryan} \emph {et~al.}}]{Khachatryan:2015bnx}%
  \BibitemOpen
  \bibfield  {author} {\bibinfo {author} {\bibfnamefont {V.}~\bibnamefont
  {Khachatryan}} \emph {et~al.} (\bibinfo {collaboration} {CMS}),\ }\href
  {\doibase 10.1103/PhysRevD.92.032008} {\bibfield  {journal} {\bibinfo
  {journal} {Phys. Rev.}\ }\textbf {\bibinfo {volume} {D92}},\ \bibinfo {pages}
  {032008} (\bibinfo {year} {2015})},\ \Eprint
  {http://arxiv.org/abs/1506.01010} {arXiv:1506.01010 [hep-ex]} \BibitemShut
  {NoStop}%
\bibitem [{\citenamefont {Zeppenfeld}\ \emph {et~al.}(2000)\citenamefont
  {Zeppenfeld}, \citenamefont {Kinnunen}, \citenamefont {Nikitenko},\ and\
  \citenamefont {Richter-Was}}]{Zeppenfeld:2000td}%
  \BibitemOpen
  \bibfield  {author} {\bibinfo {author} {\bibfnamefont {D.}~\bibnamefont
  {Zeppenfeld}}, \bibinfo {author} {\bibfnamefont {R.}~\bibnamefont
  {Kinnunen}}, \bibinfo {author} {\bibfnamefont {A.}~\bibnamefont {Nikitenko}},
  \ and\ \bibinfo {author} {\bibfnamefont {E.}~\bibnamefont {Richter-Was}},\
  }\href {\doibase 10.1103/PhysRevD.62.013009} {\bibfield  {journal} {\bibinfo
  {journal} {Phys. Rev.}\ }\textbf {\bibinfo {volume} {D62}},\ \bibinfo {pages}
  {013009} (\bibinfo {year} {2000})},\ \Eprint
  {http://arxiv.org/abs/hep-ph/0002036} {arXiv:hep-ph/0002036 [hep-ph]}
  \BibitemShut {NoStop}%
\bibitem [{\citenamefont {ATLAS}\ and\ \citenamefont
  {CMS}(2015)}]{ATLASCMS:2015higgs}%
  \BibitemOpen
  \bibfield  {author} {\bibinfo {author} {\bibnamefont {ATLAS}}\ and\ \bibinfo
  {author} {\bibnamefont {CMS}},\ }\href@noop {} {\  (\bibinfo {year}
  {2015})},\ \bibinfo {note}
  {\href{http://cds.cern.ch/record/2052552}{ATLAS-CONF-2015-044}}\BibitemShut
  {NoStop}%
\bibitem [{\citenamefont {Bolzoni}\ \emph {et~al.}(2010)\citenamefont
  {Bolzoni}, \citenamefont {Maltoni}, \citenamefont {Moch},\ and\ \citenamefont
  {Zaro}}]{Bolzoni:2010xr}%
  \BibitemOpen
  \bibfield  {author} {\bibinfo {author} {\bibfnamefont {P.}~\bibnamefont
  {Bolzoni}}, \bibinfo {author} {\bibfnamefont {F.}~\bibnamefont {Maltoni}},
  \bibinfo {author} {\bibfnamefont {S.-O.}\ \bibnamefont {Moch}}, \ and\
  \bibinfo {author} {\bibfnamefont {M.}~\bibnamefont {Zaro}},\ }\href {\doibase
  10.1103/PhysRevLett.105.011801} {\bibfield  {journal} {\bibinfo  {journal}
  {Phys. Rev. Lett.}\ }\textbf {\bibinfo {volume} {105}},\ \bibinfo {pages}
  {011801} (\bibinfo {year} {2010})},\ \Eprint {http://arxiv.org/abs/1003.4451}
  {arXiv:1003.4451 [hep-ph]} \BibitemShut {NoStop}%
\bibitem [{\citenamefont {Bolzoni}\ \emph {et~al.}(2012)\citenamefont
  {Bolzoni}, \citenamefont {Maltoni}, \citenamefont {Moch},\ and\ \citenamefont
  {Zaro}}]{Bolzoni:2011cu}%
  \BibitemOpen
  \bibfield  {author} {\bibinfo {author} {\bibfnamefont {P.}~\bibnamefont
  {Bolzoni}}, \bibinfo {author} {\bibfnamefont {F.}~\bibnamefont {Maltoni}},
  \bibinfo {author} {\bibfnamefont {S.-O.}\ \bibnamefont {Moch}}, \ and\
  \bibinfo {author} {\bibfnamefont {M.}~\bibnamefont {Zaro}},\ }\href {\doibase
  10.1103/PhysRevD.85.035002} {\bibfield  {journal} {\bibinfo  {journal} {Phys.
  Rev.}\ }\textbf {\bibinfo {volume} {D85}},\ \bibinfo {pages} {035002}
  (\bibinfo {year} {2012})},\ \Eprint {http://arxiv.org/abs/1109.3717}
  {arXiv:1109.3717 [hep-ph]} \BibitemShut {NoStop}%
\bibitem [{\citenamefont {Han}\ \emph {et~al.}(1992)\citenamefont {Han},
  \citenamefont {Valencia},\ and\ \citenamefont {Willenbrock}}]{Han:1992hr}%
  \BibitemOpen
  \bibfield  {author} {\bibinfo {author} {\bibfnamefont {T.}~\bibnamefont
  {Han}}, \bibinfo {author} {\bibfnamefont {G.}~\bibnamefont {Valencia}}, \
  and\ \bibinfo {author} {\bibfnamefont {S.}~\bibnamefont {Willenbrock}},\
  }\href {\doibase 10.1103/PhysRevLett.69.3274} {\bibfield  {journal} {\bibinfo
   {journal} {Phys. Rev. Lett.}\ }\textbf {\bibinfo {volume} {69}},\ \bibinfo
  {pages} {3274} (\bibinfo {year} {1992})},\ \Eprint
  {http://arxiv.org/abs/hep-ph/9206246} {arXiv:hep-ph/9206246 [hep-ph]}
  \BibitemShut {NoStop}%
\bibitem [{\citenamefont {Cacciari}\ \emph {et~al.}(2015)\citenamefont
  {Cacciari}, \citenamefont {Dreyer}, \citenamefont {Karlberg}, \citenamefont
  {Salam},\ and\ \citenamefont {Zanderighi}}]{Cacciari:2015jma}%
  \BibitemOpen
  \bibfield  {author} {\bibinfo {author} {\bibfnamefont {M.}~\bibnamefont
  {Cacciari}}, \bibinfo {author} {\bibfnamefont {F.~A.}\ \bibnamefont
  {Dreyer}}, \bibinfo {author} {\bibfnamefont {A.}~\bibnamefont {Karlberg}},
  \bibinfo {author} {\bibfnamefont {G.~P.}\ \bibnamefont {Salam}}, \ and\
  \bibinfo {author} {\bibfnamefont {G.}~\bibnamefont {Zanderighi}},\ }\href
  {\doibase 10.1103/PhysRevLett.115.082002} {\bibfield  {journal} {\bibinfo
  {journal} {Phys. Rev. Lett.}\ }\textbf {\bibinfo {volume} {115}},\ \bibinfo
  {pages} {082002} (\bibinfo {year} {2015})},\ \Eprint
  {http://arxiv.org/abs/1506.02660} {arXiv:1506.02660 [hep-ph]} \BibitemShut
  {NoStop}%
\bibitem [{\citenamefont {Ciccolini}\ \emph {et~al.}(2008)\citenamefont
  {Ciccolini}, \citenamefont {Denner},\ and\ \citenamefont
  {Dittmaier}}]{Ciccolini:2007ec}%
  \BibitemOpen
  \bibfield  {author} {\bibinfo {author} {\bibfnamefont {M.}~\bibnamefont
  {Ciccolini}}, \bibinfo {author} {\bibfnamefont {A.}~\bibnamefont {Denner}}, \
  and\ \bibinfo {author} {\bibfnamefont {S.}~\bibnamefont {Dittmaier}},\ }\href
  {\doibase 10.1103/PhysRevD.77.013002} {\bibfield  {journal} {\bibinfo
  {journal} {Phys. Rev.}\ }\textbf {\bibinfo {volume} {D77}},\ \bibinfo {pages}
  {013002} (\bibinfo {year} {2008})},\ \Eprint {http://arxiv.org/abs/0710.4749}
  {arXiv:0710.4749 [hep-ph]} \BibitemShut {NoStop}%
\bibitem [{\citenamefont {Harlander}\ \emph {et~al.}(2008)\citenamefont
  {Harlander}, \citenamefont {Vollinga},\ and\ \citenamefont
  {Weber}}]{Harlander:2008xn}%
  \BibitemOpen
  \bibfield  {author} {\bibinfo {author} {\bibfnamefont {R.~V.}\ \bibnamefont
  {Harlander}}, \bibinfo {author} {\bibfnamefont {J.}~\bibnamefont {Vollinga}},
  \ and\ \bibinfo {author} {\bibfnamefont {M.~M.}\ \bibnamefont {Weber}},\
  }\href {\doibase 10.1103/PhysRevD.77.053010} {\bibfield  {journal} {\bibinfo
  {journal} {Phys. Rev.}\ }\textbf {\bibinfo {volume} {D77}},\ \bibinfo {pages}
  {053010} (\bibinfo {year} {2008})},\ \Eprint {http://arxiv.org/abs/0801.3355}
  {arXiv:0801.3355 [hep-ph]} \BibitemShut {NoStop}%
\bibitem [{\citenamefont {Anastasiou}\ \emph {et~al.}(2015)\citenamefont
  {Anastasiou}, \citenamefont {Duhr}, \citenamefont {Dulat}, \citenamefont
  {Herzog},\ and\ \citenamefont {Mistlberger}}]{Anastasiou:2015ema}%
  \BibitemOpen
  \bibfield  {author} {\bibinfo {author} {\bibfnamefont {C.}~\bibnamefont
  {Anastasiou}}, \bibinfo {author} {\bibfnamefont {C.}~\bibnamefont {Duhr}},
  \bibinfo {author} {\bibfnamefont {F.}~\bibnamefont {Dulat}}, \bibinfo
  {author} {\bibfnamefont {F.}~\bibnamefont {Herzog}}, \ and\ \bibinfo {author}
  {\bibfnamefont {B.}~\bibnamefont {Mistlberger}},\ }\href {\doibase
  10.1103/PhysRevLett.114.212001} {\bibfield  {journal} {\bibinfo  {journal}
  {Phys. Rev. Lett.}\ }\textbf {\bibinfo {volume} {114}},\ \bibinfo {pages}
  {212001} (\bibinfo {year} {2015})},\ \Eprint
  {http://arxiv.org/abs/1503.06056} {arXiv:1503.06056 [hep-ph]} \BibitemShut
  {NoStop}%
\bibitem [{\citenamefont {Moch}\ \emph {et~al.}(2008)\citenamefont {Moch},
  \citenamefont {Rogal},\ and\ \citenamefont {Vogt}}]{Moch:2007rq}%
  \BibitemOpen
  \bibfield  {author} {\bibinfo {author} {\bibfnamefont {S.}~\bibnamefont
  {Moch}}, \bibinfo {author} {\bibfnamefont {M.}~\bibnamefont {Rogal}}, \ and\
  \bibinfo {author} {\bibfnamefont {A.}~\bibnamefont {Vogt}},\ }\href {\doibase
  10.1016/j.nuclphysb.2007.09.022} {\bibfield  {journal} {\bibinfo  {journal}
  {Nucl. Phys.}\ }\textbf {\bibinfo {volume} {B790}},\ \bibinfo {pages} {317}
  (\bibinfo {year} {2008})},\ \Eprint {http://arxiv.org/abs/0708.3731}
  {arXiv:0708.3731 [hep-ph]} \BibitemShut {NoStop}%
\bibitem [{\citenamefont {Vermaseren}\ \emph {et~al.}(2005)\citenamefont
  {Vermaseren}, \citenamefont {Vogt},\ and\ \citenamefont
  {Moch}}]{Vermaseren:2005qc}%
  \BibitemOpen
  \bibfield  {author} {\bibinfo {author} {\bibfnamefont {J.~A.~M.}\
  \bibnamefont {Vermaseren}}, \bibinfo {author} {\bibfnamefont
  {A.}~\bibnamefont {Vogt}}, \ and\ \bibinfo {author} {\bibfnamefont
  {S.}~\bibnamefont {Moch}},\ }\href {\doibase 10.1016/j.nuclphysb.2005.06.020}
  {\bibfield  {journal} {\bibinfo  {journal} {Nucl. Phys.}\ }\textbf {\bibinfo
  {volume} {B724}},\ \bibinfo {pages} {3} (\bibinfo {year} {2005})},\ \Eprint
  {http://arxiv.org/abs/hep-ph/0504242} {arXiv:hep-ph/0504242 [hep-ph]}
  \BibitemShut {NoStop}%
\bibitem [{\citenamefont {Furmanski}\ and\ \citenamefont
  {Petronzio}(1982)}]{Furmanski:1981cw}%
  \BibitemOpen
  \bibfield  {author} {\bibinfo {author} {\bibfnamefont {W.}~\bibnamefont
  {Furmanski}}\ and\ \bibinfo {author} {\bibfnamefont {R.}~\bibnamefont
  {Petronzio}},\ }\href {\doibase 10.1007/BF01578280} {\bibfield  {journal}
  {\bibinfo  {journal} {Z. Phys.}\ }\textbf {\bibinfo {volume} {C11}},\
  \bibinfo {pages} {293} (\bibinfo {year} {1982})}\BibitemShut {NoStop}%
\bibitem [{\citenamefont {van Neerven}\ and\ \citenamefont
  {Vogt}(2000{\natexlab{a}})}]{vanNeerven:2000uj}%
  \BibitemOpen
  \bibfield  {author} {\bibinfo {author} {\bibfnamefont {W.~L.}\ \bibnamefont
  {van Neerven}}\ and\ \bibinfo {author} {\bibfnamefont {A.}~\bibnamefont
  {Vogt}},\ }\href {\doibase 10.1016/S0550-3213(00)00480-6} {\bibfield
  {journal} {\bibinfo  {journal} {Nucl. Phys.}\ }\textbf {\bibinfo {volume}
  {B588}},\ \bibinfo {pages} {345} (\bibinfo {year} {2000}{\natexlab{a}})},\
  \Eprint {http://arxiv.org/abs/hep-ph/0006154} {arXiv:hep-ph/0006154 [hep-ph]}
  \BibitemShut {NoStop}%
\bibitem [{\citenamefont {Buehler}\ and\ \citenamefont
  {Lazopoulos}(2013)}]{Buehler:2013fha}%
  \BibitemOpen
  \bibfield  {author} {\bibinfo {author} {\bibfnamefont {S.}~\bibnamefont
  {Buehler}}\ and\ \bibinfo {author} {\bibfnamefont {A.}~\bibnamefont
  {Lazopoulos}},\ }\href {\doibase 10.1007/JHEP10(2013)096} {\bibfield
  {journal} {\bibinfo  {journal} {JHEP}\ }\textbf {\bibinfo {volume} {10}},\
  \bibinfo {pages} {096} (\bibinfo {year} {2013})},\ \Eprint
  {http://arxiv.org/abs/1306.2223} {arXiv:1306.2223 [hep-ph]} \BibitemShut
  {NoStop}%
\bibitem [{\citenamefont {Dreyer}\ and\ \citenamefont
  {Karlberg}()}]{FDAKadditional}%
  \BibitemOpen
  \bibfield  {author} {\bibinfo {author} {\bibfnamefont {F.~A.}\ \bibnamefont
  {Dreyer}}\ and\ \bibinfo {author} {\bibfnamefont {A.}~\bibnamefont
  {Karlberg}},\ }\href@noop {} {\enquote {\bibinfo {title} {{supplemental
  material, available at the end of the arXiv version of this article.}}}\
  }\BibitemShut {NoStop}%
\bibitem [{\citenamefont {Anastasiou}\ \emph {et~al.}(2016)\citenamefont
  {Anastasiou}, \citenamefont {Duhr}, \citenamefont {Dulat}, \citenamefont
  {Furlan}, \citenamefont {Gehrmann}, \citenamefont {Herzog}, \citenamefont
  {Lazopoulos},\ and\ \citenamefont {Mistlberger}}]{Anastasiou:2016cez}%
  \BibitemOpen
  \bibfield  {author} {\bibinfo {author} {\bibfnamefont {C.}~\bibnamefont
  {Anastasiou}}, \bibinfo {author} {\bibfnamefont {C.}~\bibnamefont {Duhr}},
  \bibinfo {author} {\bibfnamefont {F.}~\bibnamefont {Dulat}}, \bibinfo
  {author} {\bibfnamefont {E.}~\bibnamefont {Furlan}}, \bibinfo {author}
  {\bibfnamefont {T.}~\bibnamefont {Gehrmann}}, \bibinfo {author}
  {\bibfnamefont {F.}~\bibnamefont {Herzog}}, \bibinfo {author} {\bibfnamefont
  {A.}~\bibnamefont {Lazopoulos}}, \ and\ \bibinfo {author} {\bibfnamefont
  {B.}~\bibnamefont {Mistlberger}},\ }\href@noop {} {\  (\bibinfo {year}
  {2016})},\ \Eprint {http://arxiv.org/abs/1602.00695} {arXiv:1602.00695
  [hep-ph]} \BibitemShut {NoStop}%
\bibitem [{\citenamefont {Forte}\ \emph {et~al.}(2014)\citenamefont {Forte},
  \citenamefont {Isgrò},\ and\ \citenamefont {Vita}}]{Forte:2013mda}%
  \BibitemOpen
  \bibfield  {author} {\bibinfo {author} {\bibfnamefont {S.}~\bibnamefont
  {Forte}}, \bibinfo {author} {\bibfnamefont {A.}~\bibnamefont {Isgrò}}, \
  and\ \bibinfo {author} {\bibfnamefont {G.}~\bibnamefont {Vita}},\ }\href
  {\doibase 10.1016/j.physletb.2014.02.027} {\bibfield  {journal} {\bibinfo
  {journal} {Phys. Lett.}\ }\textbf {\bibinfo {volume} {B731}},\ \bibinfo
  {pages} {136} (\bibinfo {year} {2014})},\ \Eprint
  {http://arxiv.org/abs/1312.6688} {arXiv:1312.6688 [hep-ph]} \BibitemShut
  {NoStop}%
\bibitem [{\citenamefont {Butterworth}\ \emph {et~al.}(2016)\citenamefont
  {Butterworth} \emph {et~al.}}]{Butterworth:2015oua}%
  \BibitemOpen
  \bibfield  {author} {\bibinfo {author} {\bibfnamefont {J.}~\bibnamefont
  {Butterworth}} \emph {et~al.},\ }\href {\doibase
  10.1088/0954-3899/43/2/023001} {\bibfield  {journal} {\bibinfo  {journal} {J.
  Phys.}\ }\textbf {\bibinfo {volume} {G43}},\ \bibinfo {pages} {023001}
  (\bibinfo {year} {2016})},\ \Eprint {http://arxiv.org/abs/1510.03865}
  {arXiv:1510.03865 [hep-ph]} \BibitemShut {NoStop}%
\bibitem [{\citenamefont {Nason}\ and\ \citenamefont
  {Oleari}(2010)}]{Nason:2009ai}%
  \BibitemOpen
  \bibfield  {author} {\bibinfo {author} {\bibfnamefont {P.}~\bibnamefont
  {Nason}}\ and\ \bibinfo {author} {\bibfnamefont {C.}~\bibnamefont {Oleari}},\
  }\href {\doibase 10.1007/JHEP02(2010)037} {\bibfield  {journal} {\bibinfo
  {journal} {JHEP}\ }\textbf {\bibinfo {volume} {02}},\ \bibinfo {pages} {037}
  (\bibinfo {year} {2010})},\ \Eprint {http://arxiv.org/abs/0911.5299}
  {arXiv:0911.5299 [hep-ph]} \BibitemShut {NoStop}%
\bibitem [{\citenamefont {Sanchez~Guillen}\ \emph {et~al.}(1991)\citenamefont
  {Sanchez~Guillen}, \citenamefont {Miramontes}, \citenamefont {Miramontes},
  \citenamefont {Parente},\ and\ \citenamefont
  {Sampayo}}]{SanchezGuillen:1990iq}%
  \BibitemOpen
  \bibfield  {author} {\bibinfo {author} {\bibfnamefont {J.}~\bibnamefont
  {Sanchez~Guillen}}, \bibinfo {author} {\bibfnamefont {J.}~\bibnamefont
  {Miramontes}}, \bibinfo {author} {\bibfnamefont {M.}~\bibnamefont
  {Miramontes}}, \bibinfo {author} {\bibfnamefont {G.}~\bibnamefont {Parente}},
  \ and\ \bibinfo {author} {\bibfnamefont {O.~A.}\ \bibnamefont {Sampayo}},\
  }\href {\doibase 10.1016/0550-3213(91)90340-4} {\bibfield  {journal}
  {\bibinfo  {journal} {Nucl. Phys.}\ }\textbf {\bibinfo {volume} {B353}},\
  \bibinfo {pages} {337} (\bibinfo {year} {1991})}\BibitemShut {NoStop}%
\bibitem [{\citenamefont {van Neerven}\ and\ \citenamefont
  {Zijlstra}(1991)}]{vanNeerven:1991nn}%
  \BibitemOpen
  \bibfield  {author} {\bibinfo {author} {\bibfnamefont {W.~L.}\ \bibnamefont
  {van Neerven}}\ and\ \bibinfo {author} {\bibfnamefont {E.~B.}\ \bibnamefont
  {Zijlstra}},\ }\href {\doibase 10.1016/0370-2693(91)91024-P} {\bibfield
  {journal} {\bibinfo  {journal} {Phys. Lett.}\ }\textbf {\bibinfo {volume}
  {B272}},\ \bibinfo {pages} {127} (\bibinfo {year} {1991})}\BibitemShut
  {NoStop}%
\bibitem [{\citenamefont {Zijlstra}\ and\ \citenamefont {van
  Neerven}(1992{\natexlab{a}})}]{Zijlstra:1992qd}%
  \BibitemOpen
  \bibfield  {author} {\bibinfo {author} {\bibfnamefont {E.~B.}\ \bibnamefont
  {Zijlstra}}\ and\ \bibinfo {author} {\bibfnamefont {W.~L.}\ \bibnamefont {van
  Neerven}},\ }\href {\doibase 10.1016/0550-3213(92)90087-R} {\bibfield
  {journal} {\bibinfo  {journal} {Nucl. Phys.}\ }\textbf {\bibinfo {volume}
  {B383}},\ \bibinfo {pages} {525} (\bibinfo {year}
  {1992}{\natexlab{a}})}\BibitemShut {NoStop}%
\bibitem [{\citenamefont {Zijlstra}\ and\ \citenamefont {van
  Neerven}(1992{\natexlab{b}})}]{Zijlstra:1992kj}%
  \BibitemOpen
  \bibfield  {author} {\bibinfo {author} {\bibfnamefont {E.~B.}\ \bibnamefont
  {Zijlstra}}\ and\ \bibinfo {author} {\bibfnamefont {W.~L.}\ \bibnamefont {van
  Neerven}},\ }\href {\doibase 10.1016/0370-2693(92)91277-G} {\bibfield
  {journal} {\bibinfo  {journal} {Phys. Lett.}\ }\textbf {\bibinfo {volume}
  {B297}},\ \bibinfo {pages} {377} (\bibinfo {year}
  {1992}{\natexlab{b}})}\BibitemShut {NoStop}%
\bibitem [{\citenamefont {van Neerven}\ and\ \citenamefont
  {Vogt}(2000{\natexlab{b}})}]{vanNeerven:1999ca}%
  \BibitemOpen
  \bibfield  {author} {\bibinfo {author} {\bibfnamefont {W.~L.}\ \bibnamefont
  {van Neerven}}\ and\ \bibinfo {author} {\bibfnamefont {A.}~\bibnamefont
  {Vogt}},\ }\href {\doibase 10.1016/S0550-3213(99)00668-9} {\bibfield
  {journal} {\bibinfo  {journal} {Nucl. Phys.}\ }\textbf {\bibinfo {volume}
  {B568}},\ \bibinfo {pages} {263} (\bibinfo {year} {2000}{\natexlab{b}})},\
  \Eprint {http://arxiv.org/abs/hep-ph/9907472} {arXiv:hep-ph/9907472 [hep-ph]}
  \BibitemShut {NoStop}%
\bibitem [{\citenamefont {Moch}\ \emph {et~al.}(2005)\citenamefont {Moch},
  \citenamefont {Vermaseren},\ and\ \citenamefont {Vogt}}]{Moch:2004xu}%
  \BibitemOpen
  \bibfield  {author} {\bibinfo {author} {\bibfnamefont {S.}~\bibnamefont
  {Moch}}, \bibinfo {author} {\bibfnamefont {J.~A.~M.}\ \bibnamefont
  {Vermaseren}}, \ and\ \bibinfo {author} {\bibfnamefont {A.}~\bibnamefont
  {Vogt}},\ }\href {\doibase 10.1016/j.physletb.2004.11.063} {\bibfield
  {journal} {\bibinfo  {journal} {Phys. Lett.}\ }\textbf {\bibinfo {volume}
  {B606}},\ \bibinfo {pages} {123} (\bibinfo {year} {2005})},\ \Eprint
  {http://arxiv.org/abs/hep-ph/0411112} {arXiv:hep-ph/0411112 [hep-ph]}
  \BibitemShut {NoStop}%
\bibitem [{\citenamefont {Vogt}\ \emph {et~al.}(2006)\citenamefont {Vogt},
  \citenamefont {Moch},\ and\ \citenamefont {Vermaseren}}]{Vogt:2006bt}%
  \BibitemOpen
  \bibfield  {author} {\bibinfo {author} {\bibfnamefont {A.}~\bibnamefont
  {Vogt}}, \bibinfo {author} {\bibfnamefont {S.}~\bibnamefont {Moch}}, \ and\
  \bibinfo {author} {\bibfnamefont {J.}~\bibnamefont {Vermaseren}},\ }\bibfield
   {booktitle} {\emph {\bibinfo {booktitle} {{Proceedings, 8th DESY Workshop on
  Elementary Particle Theory: Loops and Legs in Quantum Field Theory}}},\
  }\href {\doibase 10.1016/j.nuclphysbps.2006.09.101} {\bibfield  {journal}
  {\bibinfo  {journal} {Nucl. Phys. Proc. Suppl.}\ }\textbf {\bibinfo {volume}
  {160}},\ \bibinfo {pages} {44} (\bibinfo {year} {2006})},\ \bibinfo {note}
  {[,44(2006)]},\ \Eprint {http://arxiv.org/abs/hep-ph/0608307}
  {arXiv:hep-ph/0608307 [hep-ph]} \BibitemShut {NoStop}%
\bibitem [{\citenamefont {Salam}\ and\ \citenamefont
  {Rojo}(2009)}]{Salam:2008qg}%
  \BibitemOpen
  \bibfield  {author} {\bibinfo {author} {\bibfnamefont {G.~P.}\ \bibnamefont
  {Salam}}\ and\ \bibinfo {author} {\bibfnamefont {J.}~\bibnamefont {Rojo}},\
  }\href {\doibase 10.1016/j.cpc.2008.08.010} {\bibfield  {journal} {\bibinfo
  {journal} {Comput. Phys. Commun.}\ }\textbf {\bibinfo {volume} {180}},\
  \bibinfo {pages} {120} (\bibinfo {year} {2009})},\ \Eprint
  {http://arxiv.org/abs/0804.3755} {arXiv:0804.3755 [hep-ph]} \BibitemShut
  {NoStop}%
\bibitem [{\citenamefont {Dulat}\ \emph {et~al.}(2016)\citenamefont {Dulat},
  \citenamefont {Hou}, \citenamefont {Gao}, \citenamefont {Guzzi},
  \citenamefont {Huston}, \citenamefont {Nadolsky}, \citenamefont {Pumplin},
  \citenamefont {Schmidt}, \citenamefont {Stump},\ and\ \citenamefont
  {Yuan}}]{Dulat:2015mca}%
  \BibitemOpen
  \bibfield  {author} {\bibinfo {author} {\bibfnamefont {S.}~\bibnamefont
  {Dulat}}, \bibinfo {author} {\bibfnamefont {T.-J.}\ \bibnamefont {Hou}},
  \bibinfo {author} {\bibfnamefont {J.}~\bibnamefont {Gao}}, \bibinfo {author}
  {\bibfnamefont {M.}~\bibnamefont {Guzzi}}, \bibinfo {author} {\bibfnamefont
  {J.}~\bibnamefont {Huston}}, \bibinfo {author} {\bibfnamefont
  {P.}~\bibnamefont {Nadolsky}}, \bibinfo {author} {\bibfnamefont
  {J.}~\bibnamefont {Pumplin}}, \bibinfo {author} {\bibfnamefont
  {C.}~\bibnamefont {Schmidt}}, \bibinfo {author} {\bibfnamefont
  {D.}~\bibnamefont {Stump}}, \ and\ \bibinfo {author} {\bibfnamefont {C.~P.}\
  \bibnamefont {Yuan}},\ }\href {\doibase 10.1103/PhysRevD.93.033006}
  {\bibfield  {journal} {\bibinfo  {journal} {Phys. Rev.}\ }\textbf {\bibinfo
  {volume} {D93}},\ \bibinfo {pages} {033006} (\bibinfo {year} {2016})},\
  \Eprint {http://arxiv.org/abs/1506.07443} {arXiv:1506.07443 [hep-ph]}
  \BibitemShut {NoStop}%
\bibitem [{\citenamefont {Harland-Lang}\ \emph {et~al.}(2015)\citenamefont
  {Harland-Lang}, \citenamefont {Martin}, \citenamefont {Motylinski},\ and\
  \citenamefont {Thorne}}]{Harland-Lang:2014zoa}%
  \BibitemOpen
  \bibfield  {author} {\bibinfo {author} {\bibfnamefont {L.~A.}\ \bibnamefont
  {Harland-Lang}}, \bibinfo {author} {\bibfnamefont {A.~D.}\ \bibnamefont
  {Martin}}, \bibinfo {author} {\bibfnamefont {P.}~\bibnamefont {Motylinski}},
  \ and\ \bibinfo {author} {\bibfnamefont {R.~S.}\ \bibnamefont {Thorne}},\
  }\href {\doibase 10.1140/epjc/s10052-015-3397-6} {\bibfield  {journal}
  {\bibinfo  {journal} {Eur. Phys. J.}\ }\textbf {\bibinfo {volume} {C75}},\
  \bibinfo {pages} {204} (\bibinfo {year} {2015})},\ \Eprint
  {http://arxiv.org/abs/1412.3989} {arXiv:1412.3989 [hep-ph]} \BibitemShut
  {NoStop}%
\bibitem [{\citenamefont {Ball}\ \emph {et~al.}(2015)\citenamefont {Ball} \emph
  {et~al.}}]{Ball:2014uwa}%
  \BibitemOpen
  \bibfield  {author} {\bibinfo {author} {\bibfnamefont {R.~D.}\ \bibnamefont
  {Ball}} \emph {et~al.} (\bibinfo {collaboration} {NNPDF}),\ }\href {\doibase
  10.1007/JHEP04(2015)040} {\bibfield  {journal} {\bibinfo  {journal} {JHEP}\
  }\textbf {\bibinfo {volume} {04}},\ \bibinfo {pages} {040} (\bibinfo {year}
  {2015})},\ \Eprint {http://arxiv.org/abs/1410.8849} {arXiv:1410.8849
  [hep-ph]} \BibitemShut {NoStop}%
\bibitem [{\citenamefont {Aad}\ \emph {et~al.}(2015)\citenamefont {Aad} \emph
  {et~al.}}]{Aad:2015zhl}%
  \BibitemOpen
  \bibfield  {author} {\bibinfo {author} {\bibfnamefont {G.}~\bibnamefont
  {Aad}} \emph {et~al.} (\bibinfo {collaboration} {ATLAS, CMS}),\ }\bibfield
  {booktitle} {\emph {\bibinfo {booktitle} {{Proceedings, Meeting of the APS
  Division of Particles and Fields (DPF 2015)}}},\ }\href {\doibase
  10.1103/PhysRevLett.114.191803} {\bibfield  {journal} {\bibinfo  {journal}
  {Phys. Rev. Lett.}\ }\textbf {\bibinfo {volume} {114}},\ \bibinfo {pages}
  {191803} (\bibinfo {year} {2015})},\ \Eprint
  {http://arxiv.org/abs/1503.07589} {arXiv:1503.07589 [hep-ex]} \BibitemShut
  {NoStop}%
\bibitem [{\citenamefont {Olive}\ \emph {et~al.}(2014)\citenamefont {Olive}
  \emph {et~al.}}]{Agashe:2014kda}%
  \BibitemOpen
  \bibfield  {author} {\bibinfo {author} {\bibfnamefont {K.~A.}\ \bibnamefont
  {Olive}} \emph {et~al.} (\bibinfo {collaboration} {Particle Data Group}),\
  }\href {\doibase 10.1088/1674-1137/38/9/090001} {\bibfield  {journal}
  {\bibinfo  {journal} {Chin. Phys.}\ }\textbf {\bibinfo {volume} {C38}},\
  \bibinfo {pages} {090001} (\bibinfo {year} {2014})}\BibitemShut {NoStop}%
\bibitem [{\citenamefont {Andersen}\ \emph {et~al.}(2008)\citenamefont
  {Andersen}, \citenamefont {Binoth}, \citenamefont {Heinrich},\ and\
  \citenamefont {Smillie}}]{Andersen:2007mp}%
  \BibitemOpen
  \bibfield  {author} {\bibinfo {author} {\bibfnamefont {J.~R.}\ \bibnamefont
  {Andersen}}, \bibinfo {author} {\bibfnamefont {T.}~\bibnamefont {Binoth}},
  \bibinfo {author} {\bibfnamefont {G.}~\bibnamefont {Heinrich}}, \ and\
  \bibinfo {author} {\bibfnamefont {J.~M.}\ \bibnamefont {Smillie}},\ }\href
  {\doibase 10.1088/1126-6708/2008/02/057} {\bibfield  {journal} {\bibinfo
  {journal} {JHEP}\ }\textbf {\bibinfo {volume} {02}},\ \bibinfo {pages} {057}
  (\bibinfo {year} {2008})},\ \Eprint {http://arxiv.org/abs/0709.3513}
  {arXiv:0709.3513 [hep-ph]} \BibitemShut {NoStop}%
\bibitem [{\citenamefont {proVBFH}()}]{proVBFH}%
  \BibitemOpen
  \bibfield  {author} {\bibinfo {author} {\bibnamefont {proVBFH}},\ }\href@noop
  {} {}\bibinfo {note} {{\url{http://provbfh.hepforge.org/}}}\BibitemShut
  {NoStop}%
\bibitem [{\citenamefont {Gehrmann}\ and\ \citenamefont
  {Glover}(2009)}]{Gehrmann:2009vu}%
  \BibitemOpen
  \bibfield  {author} {\bibinfo {author} {\bibfnamefont {T.}~\bibnamefont
  {Gehrmann}}\ and\ \bibinfo {author} {\bibfnamefont {E.~W.~N.}\ \bibnamefont
  {Glover}},\ }\href {\doibase 10.1016/j.physletb.2009.04.083} {\bibfield
  {journal} {\bibinfo  {journal} {Phys. Lett.}\ }\textbf {\bibinfo {volume}
  {B676}},\ \bibinfo {pages} {146} (\bibinfo {year} {2009})},\ \Eprint
  {http://arxiv.org/abs/0904.2665} {arXiv:0904.2665 [hep-ph]} \BibitemShut
  {NoStop}%
\bibitem [{\citenamefont {Vogt}\ \emph {et~al.}(2004)\citenamefont {Vogt},
  \citenamefont {Moch},\ and\ \citenamefont {Vermaseren}}]{Vogt:2004mw}%
  \BibitemOpen
  \bibfield  {author} {\bibinfo {author} {\bibfnamefont {A.}~\bibnamefont
  {Vogt}}, \bibinfo {author} {\bibfnamefont {S.}~\bibnamefont {Moch}}, \ and\
  \bibinfo {author} {\bibfnamefont {J.~A.~M.}\ \bibnamefont {Vermaseren}},\
  }\href {\doibase 10.1016/j.nuclphysb.2004.04.024} {\bibfield  {journal}
  {\bibinfo  {journal} {Nucl. Phys.}\ }\textbf {\bibinfo {volume} {B691}},\
  \bibinfo {pages} {129} (\bibinfo {year} {2004})},\ \Eprint
  {http://arxiv.org/abs/hep-ph/0404111} {arXiv:hep-ph/0404111 [hep-ph]}
  \BibitemShut {NoStop}%
\bibitem [{\citenamefont {Moch}\ \emph {et~al.}(2004)\citenamefont {Moch},
  \citenamefont {Vermaseren},\ and\ \citenamefont {Vogt}}]{Moch:2004pa}%
  \BibitemOpen
  \bibfield  {author} {\bibinfo {author} {\bibfnamefont {S.}~\bibnamefont
  {Moch}}, \bibinfo {author} {\bibfnamefont {J.}~\bibnamefont {Vermaseren}}, \
  and\ \bibinfo {author} {\bibfnamefont {A.}~\bibnamefont {Vogt}},\ }\href
  {\doibase 10.1016/j.nuclphysb.2004.03.030} {\bibfield  {journal} {\bibinfo
  {journal} {Nucl.Phys.}\ }\textbf {\bibinfo {volume} {B688}},\ \bibinfo
  {pages} {101} (\bibinfo {year} {2004})},\ \Eprint
  {http://arxiv.org/abs/hep-ph/0403192} {arXiv:hep-ph/0403192 [hep-ph]}
  \BibitemShut {NoStop}%
\end{thebibliography}%
\newpage

\onecolumngrid
\newpage
\appendix
\section*{Supplemental material}
\setcounter{page}{1}
\setcounter{figure}{0}
\section*{Supplemental material to ``Vector-boson fusion Higgs production at \NNNLO{} in QCD'' by Fr\'ed\'eric A. Dreyer and Alexander Karlberg}

\subsection{Scale variation up to \NNNLO{}}
To compute the dependence of the cross section on the values of the
factorisation and renormalisation scales, we use renormalisation
group methods~\cite{Furmanski:1981cw,vanNeerven:2000uj} on the structure functions
\begin{equation}
  \label{eq:conv-structf-suppl}
  F_i^V = \sum_a C_i^{V,a} \otimes f_a \,,\quad i=1,2,3\,.
\end{equation}
This requires us to compute the scale dependence to third order in the
coefficient functions as well as in the PDFs.

We start by evaluating the running coupling for $\as$
\begin{align}
  \label{eq:as-running}
  \as(Q) = \as(\muR) + \as^2(\muR) \beta_0 L_{RQ} 
    + \as^3(\muR) (\beta_0^2 L_{RQ}^2 + \beta_1 L_{RQ}) + \mathcal{O}(\as^4(\muR))\,,
\end{align}
where we introduced the shorthand notation
\begin{equation}
  \label{eq:LRQ-notation}
  L_{RQ} = \ln\left(\frac{\muR^2}{Q^2}\right)\,,\quad
  L_{FQ} = \ln\left(\frac{\muF^2}{Q^2}\right)\,,
\end{equation}
as well as $\beta_0 = (33 - 2 n_f)/12\pi$ and $\beta_1 = (153 - 19 n_f)/24\pi^2$. The
coefficient functions can thus be expressed as an expansion in $\as(\mu_R)$
\begin{multline}
  \label{eq:coef-fct-expansion}
  C_i^{(0)} + \frac{\as(Q)}{2\pi} C_i^{(1)} + \left(\frac{\as(Q)}{2\pi}\right)^2 C_i^{(2)}
  + \left(\frac{\as(Q)}{2\pi}\right)^3 C_i^{(3)} =  
  C_i^{(0)} + \frac{\as(\muR)}{2\pi} C_i^{(1)} +
  \left(\frac{\as(\muR)}{2\pi}\right)^2 \left(C_i^{(2)} + 2\pi \beta_0 C_i^{(1)} L_{RQ}\right)
  \\ 
  + \left(\frac{\as(\muR)}{2\pi}\right)^3 \bigg[
    C_i^{(3)} + 4\pi \beta_0 C_i^{(2)} L_{RQ}     
    + 4\pi^2 C_i^{(1)} L_{RQ} (\beta_1 + \beta_0^2 L_{RQ})
  \bigg]\,.
\end{multline}
To evaluate the dependence of the PDFs on the factorisation scale $\muF$, we
integrate the DGLAP equation, using
\begin{equation}
  \label{eq:pdf-integ}
  f(x,Q) = f(x,\muF) - \int_0^{L_{FQ}} dL \frac{d}{dL} f(x,\mu) \,.
\end{equation}
It is then straightforward to express the PDF in terms of an expansion
in $\as(\muR)$ evaluated at $\muF$.
We have
\begin{multline}
  \label{eq:pdf-expansion}
  f(x,Q) = f(x, \muF) - \frac{\as(\muR)}{2\pi} L_{FQ} f(x, \muF)
  P^{(0)}
  - \left(\frac{\as(\muR)}{2\pi}\right)^2 L_{FQ} f(x, \muF) \Big[
  P^{(1)} - \frac12 L_{FQ} (P^{(0)})^2
  -\pi\beta_0 P^{(0)} (L_{FQ} - 2 L_{RQ}) \Big]
  \\ 
  - \left(\frac{\as(\muR)}{2\pi}\right)^3 L_{FQ} f(x, \muF) \Big[
  P^{(2)} - \frac12 L_{FQ} (P^{(0)} P^{(1)} + P^{(1)} P^{(0)})
  +  \pi \beta_0 (L_{FQ} - 2 L_{RQ}) (L_{FQ} (P^{(0)})^2 - 2 P^{(1)})
  \\ 
  + \frac16 L_{FQ}^2 (P^{(0)})^3
  + 4\pi^2 \beta_0^2 P^{(0)} (L_{RQ}^2 - L_{FQ} L_{RQ} + \frac13
  L_{FQ}^2)
  - 2 \pi^2 \beta_1 P^{(0)} (L_{FQ} - 2 L_{RQ}) \Big]\,.
\end{multline}
Here we defined the expansion of the splitting functions as
\begin{equation}
  \label{eq:split-fct-expansion}
  P(z,\as) = \sum_{i=0}{\left(\frac{\as}{2\pi}\right)^{i} P^{(i)}(z) }\,,
\end{equation}
where terms up to $P^{(2)}(z)$ are
known~\cite{Vogt:2004mw,Moch:2004pa}. 
Equations~(\ref{eq:coef-fct-expansion}) and~(\ref{eq:pdf-expansion})
allow us to evaluate the convolution in
equation~(\ref{eq:conv-structf-suppl}) up to \NNNLO{} in perturbative QCD
for any choice of the renormalisation and factorisation scales.

\end{document}